\documentclass[10pt, a4paper]{amsart}
\usepackage{amsmath}
\usepackage{latexsym}
\usepackage[all]{xy}
\usepackage{color}

\include{amslatex}

\numberwithin{equation}{section}

\begin{document}

\begin{title} [Emergence of classical gravity and...]
 {Emergence of classical gravity and the objective reduction of the quantum state in deterministic models of quantum mechanics}
\end{title}

\maketitle
\begin{center}
\author{Ricardo Gallego Torrom\'e\footnote{email: rigato39@gmail.com}}
\end{center}
\bigskip
\address{Departamento de Matem\'atica, Universidade Federal de S\~ao Carlos, Rodovia Washington Lu\'is, km 235, SP, Brazil}

\begin{abstract}
 Models for deterministic quantum mechanics of Cartan-Randers type are introduced, together with the fundamental notions of the concentration of measure theory. We explain how the application of the concentration of measure to Cartan-Randers models provides a framework from which it emerge  1. The invariance under infinitesimal  diffeomorphisms of the macroscopic dynamics 2. A mechanism for reduction of the quantum state and 3. The Weak Equivalence Principle.
\end{abstract}

\section{Introduction}
 {\it Deterministic emergent quantum mechanics} denotes several new approaches to the foundations of quantum mechanics based on deterministic descriptions of an underlying level of physical reality \cite{AcostaFernandezetIsidroSantander2013, Adler, Blasone, Elze, Groessing2013, Hooft, Hooft14, Wetterich2008}.  In deterministic emergent quantum mechanics there are two physical scales. First, there is a {\it fundamental scale}, which is usually associated with the Planck scale. The degrees of freedom at this scale are deterministic. The second scale is associated with quantum scales (Standard Model scale, or atomic or even molecular scales). The main objectives of such frameworks to reproduce the mathematical formalism of quantum mechanics as an emergent description  from an underlying fundamental deterministic framework.

 It is in this context of deterministic models for quantum mechanics that the author proposed a particular type of geometric micro-statistical models \cite{Ricardo2005}. This paper describes a theoretical mechanism for quantum state reduction in the framework of such deterministic models. It turns out that in the dynamical regime where the quantum state reduction of the quantum state happens, the interaction driving the reduction has a strong resemblance with the gravitational interaction. This provides a newer look to the possibility that the dynamical reduction of the quantum state is related with the gravitational interaction.

 The structure of this paper is the following. We first describe the dynamical systems that we will consider. After this, a succinct introduction to the {\it concentration of measure phenomena}  in measure metric spaces is provided. This is a mathematical property that appears in several areas of geometry, functional analysis and probability theory. Then we apply concentration of measure to show how the reduction of the wave packet happens spontaneously in our deterministic models for quantum systems, independently of the presence of a measurement device or process. The concentration of measure shows how the weak equivalence principle emerges as well.
\section{Deterministic Cartan-Randers Models}
In the deterministic models that we consider
the physical degrees of freedom describe point particles, whose evolution determines the worldsheets that are parameterized by a {\it two dimensional time parameter} $(t,\tau)$, where $\tau$ is a time coordinate associated to a macroscopic observer and $t$ corresponds to a new parameter that enters in the dynamics. We called $t$ the {\it internal time}. These two parameters are logically independent. These worldsheets are submanifolds  of a configuration manifold which has a product structure
\begin{align}
TM\cong \prod^N_{k=1}\, TM^k_4,
\label{configurationmanifold}
\end{align}
where $\{M^k_4,\, k=\,1,...,N\}$ are $4$-manifolds diffeomorphic to a fixed 4-manifold  $M_4$ and $N>>1$. Then we have a collection of diffeomorphisms $\{\phi_k:M^k_4\to M_4,\, k=\,1,...,N\}$.
The algebra of functions $\mathcal{F}_D(T^*TM)$ on $T^*TM$ that we consider are the {\it diagonal functions}, obtained by an algebra embedding
\begin{align}
\theta:\mathcal{F}(T^*TM_4)\to \mathcal{F}(T^*TM), \quad f\mapsto (f_1,...,f_N),\,\, f_k=f,\,\,k=1,...,N.
\label{algebramorphism}
\end{align}
that are elements of $\mathcal{F}(T^*TM)$  of the form
\begin{align*}
\big((u_1,p_1),...,(u_N,p_N)\big)\mapsto \big(f(u_1,p_1),...,f(u_N,p_N)\big).
 \end{align*}
 It is relevance for our considerations that the average values of the functions $f\in \mathcal{F}_D(T^*TM)$ are  well defined, that is, independent of the diffeomorphisms $\{\phi_k:M^k_4\to M_4,\, k=\,1,...,N\}$. This can be achieved if the measure $\mu_P$ in $T^*TM$ is of the form
\begin{align}
\mu_P=\,\prod^N_{k=1} \mu(k)_P,
\end{align}
where $\mu(k)_P$ is a probability measure in $T^*TM^k_4$. Then one defines
\begin{align}
   \widetilde{\mathcal{O}}_{CM}(\mathcal{S},t,\tau)=\,\int_{T^*TM}\,\widetilde{\mathcal{O}}(k,t,\tau)\,\mu_P.
  \end{align}

The classical Hamiltonian function for our deterministic systems is of the form
\begin{align}
H(t,\tau,u,p)=\,\sum^{8N}_{n=1}\,\beta^n(u,t,\tau)p_n,
\label{randershamiltoniant}
\end{align}
where $u\in TM$, $p\in\,T^*TM$, $TM$ is the {\it configuration manifold} with $N>>1$ and constrained by the requirement that, under an underlying metric structure $\hat{\eta}$ on $T^*TM$,
\begin{align}
\|\beta\|_{\hat{\eta}}<1.
\label{constraintonbetas}
 \end{align}
 These conditions are equivalent to have bounded acceleration and speed for each fundamental degrees of freedom. The theory that we propose goes beyond these requirements and assume the existence of  universal  maximal acceleration and maximal speed. We have called these dynamical systems deterministic Cartan-Randers models (in short, DCRM). These models were called Finslerian models in  \cite{Ricardo2005}, but the new denomination is more adequate.

The two time parameters $(t,\tau)$ are used to describe a {\it double dynamics} $U_t$ and $U_\tau$.
The dynamical equations for this Hamiltonian for the $U_\tau$ dynamics are
\begin{align}
\frac{d u^i}{d\tau}=\,\beta^i(t,\tau,u),\quad \frac{d p_i}{d\tau}=\,-\sum^{8N}_{k=1}\,\frac{\partial \beta^k(t,\tau,u)}{\partial u^i}p_k, \quad i,j,k=1,...,8N.
\label{Hamiltonianequation}
\end{align}
 The $U_t$ should hold have several properties. There must consists of cycles. Each cycle start with an  {\it ergodic regime} follow  by a {\it concentration regime}, followed  by a {\it expanding regime}, followed by the next {\it ergodic regime}, etc... The concentration regime  is described by geometric flow of the geometric structures on $T^*TM$ defining the Cartan-Randers system (that is, a dual metric $g\in\Gamma T^{(2,0)}TM$ and a vector field $\beta\in \,\Gamma TM$) and must allow for equilibrium states, characterized by the $6$-dimensional hypersurface $\Sigma=\prod_{k=1}\,\Sigma_4\times {S}_3$, where $\Sigma_4$ is the unit hyperboloid and $S_3$ is the $3$-sphere. Although we do not propose a particular form of the flow here, neither in \cite{Ricardo2005}, we will be able to extract falsifiable consequences for our theory. The research of several possible $U_t$ dynamics is under current investigation.

One of the techniques highlighted by G.'t Hooft is to describe classical systems quantum mechanically, that is, using Hilbert space theory (such ideas are closely related with Koopman's approach to dynamical systems \cite{Koopman, ReedSimonI}). In Hooft's theory the standard  canonical quantization  relations are adopted and used. Thus, commutators are taken for operators at equal time $\tau$.
 The quantum operators that we will consider are quantization  of the diagonal algebra $\mathcal{F}_D(T^*TM)$. However, in DCRM time is associated with a $2$-dimensional parameter and we need to specify the commutation relations at each fixed value of the pair $(t,\tau)$.  Therefore, we adopt the following quantization rules defined as follow,
\begin{itemize}
\item The values of the position coordinates $\{x^\mu_k,\,k=1,...,N,\,\mu=1,...,4\}$ and the velocity coordinates $\{y^\mu_k,\,k=1,...,N,\,\mu=1,...,4\}$ of the fundamental degrees of freedom appear as eigenvalues of certain hermitian operators $\{\hat{x}^\mu_k,\hat{y}^\mu_k,\,k=1,...,N,\,\mu=1,...,4\}$,
    \begin{align}
    ; ber \hat{x}^\mu_k |x^\mu_l,y^\nu_l,\mu,\nu=1,...,4\rangle=\,\sum_{l}\,\delta_{kl}\,x^\mu_l\,|x^\mu_l,y^\nu_l,\mu,\nu=1,...,4\rangle,\\
    \hat{y}^\nu_k |x^\mu_l,y^\nu_l,\mu,\nu=1,...,4\rangle=\,\sum_{l}\delta_{kl}\,y^\nu_l\,|x^\mu_l,y^\nu_l,\mu,\nu=1,...,4\rangle.
    \label{positionvelocityeigenbasis}
    \end{align}

\item There are a set of Hermitian operators $\{\hat{p}_{\mu x k},\hat{p}_{\mu y k},\,k=1,...,N,\,\mu=1,...,4\}$ that generates local diffeomorphism on $TM$ along the integral curves of the local vector fields $\{\frac{\partial}{\partial x^\mu_k},\,\frac{\partial}{\partial x^\nu_k},\quad \,\mu,\nu=1,2,3,4,\,k=1,...,N\}$.

\item The canonical commutation relations at fixed $2$-time $(t,\tau)$ hold good,
\begin{align}
[\hat{x}^\mu_k,\hat{p}_{\nu xl}]= \,\imath\,\hbar\,\delta^\mu_\nu\,\delta_{kl},\,\,\quad [\hat{y}^\mu_k,\hat{p}_{\nu yl}]=\, \imath\,\hbar\,\delta^\mu_\nu\,\delta_{kl}\,
\label{quantumconditions}
\end{align}
and any other canonical commutator is equal to zero.
\end{itemize}
The Hamiltonian \eqref{randershamiltoniant} can be quantized to obtain the following Hermitian operator,
 \begin{align}
\widehat{H}(t,\tau,\hat{u},\hat{p})=\,\frac{1}{2}\sum^{8N}_{k=1}\,\big(\beta^k(t,\tau,\hat{u})\,\hat{p}_k+\,\hat{p}_k\,\beta^k(t,\tau,\hat{u})\big)
\label{quantumHamiltonian}
\end{align}
When the quantum conditions \eqref{quantumconditions} are applied to the Heisenberg's equation for the Hamiltonian \eqref{quantumHamiltonian}, the equations \eqref{Hamiltonianequation} for the eigenvalues $\{x^\mu_k,y^\mu_k\,k=1,...,N,\,\mu=1,...,4\}$ in the eigenbasis \eqref{positionvelocityeigenbasis} are obtained.
Moreover,
in order to be consistent between the Hilbert formalism and the geometric formalism, we need to impose the constraints
\begin{align}
\hat{y}^\mu_k=\,\frac{d \hat{x}^\mu_k}{d\tau},\quad \forall \,\,k=\,1,...,N,\quad \mu=1,...,4.
\label{constraintsonthevelocity}
\end{align}
These relations are consistent, despite the fact that in quantum mechanics the
coordinate operators $\{\hat{X}^a, a=2,3,4\}$ and the velocity operators $\{\dot{\hat{X}}^a,\,a=2,3,4\}$ do not commute for each quantum degree of freedom $a$. However, let us remark that the operators $\{(\hat{x}^a_k,\hat{y}^a_k)\}^{N,4}_{k=1,a=2}$ do not coincide with the quantum operators $\{{X}^a,\dot{\hat{X}}^a,\,a=2,3,4\}$. The operators $\{\hat{X}^a,\dot{\hat{X}}^a,\,a=2,3,4\}$ have as spectrum the possible outcomes of measurements of position and velocity for each $a$. The operators $\{\hat{X}^a,\dot{\hat{X}}^a,\,a=2,3,4\}$ should emerge in DCRM together with the wave function for the quantum state, that appears as phenomenological description of the ergodic regime when one considers the  projection $(t,\tau)\mapsto \tau$. It is currently under investigation to construct systematically the operators $\{\hat{X}^a,\dot{\hat{X}}^a\, a=2,3,4\}$ from the canonical operators $\{x^\mu,y^\mu,\hat{p}_{\mu x k},\hat{p}_{\mu, y k},\,\mu=1,2,3,4,\,=1,...,N\}$.

 As a consequence of this interpretation, the quantum states that are obtained from DCRM are generically non-localized in both position and speed (or canonical momentum) and  both $\hat{X}^a$ and $\dot{\hat{X}}^a$ have generally non-zero dispersion. Therefore, the quantum states  must hold a representation of a non-commutative algebra,
\begin{align}
[\hat{X}^\mu_a,\hat{X}^\nu_b]= A^{\mu\nu}\,\delta_{ab},\quad [\dot{\hat{X}}^\mu_a,\dot{\hat{X}}^\nu_b]= B^{\mu\nu}\,\delta_{ab},\quad [\hat{X}^\mu,\dot{\hat{X}}^\nu]=C^{\mu\nu}\,\delta_{ab},
\label{noncommutativealgebra}
\end{align}
with $\mu,\nu=1,2,3,4, a,b=1,...,N.$
 The requirement of local invariance under the Lorentz group of this algebra implies that the spacetime must be a quantum spacetime compatible with an universal maximal acceleration and a universal maximal speed. A geometric realization of quantum spacetime with maximal acceleration and maximal speed is under current investigation.

A fundamental problem with the quantum Hermitian  Hamiltonian operator \eqref{quantumHamiltonian} is that, being linear on the momentum operators, it is not direct that the quantum Hamiltonian $\widehat{H}(t,\tau,\hat{u},\hat{p})$ has a stable vacuum eigenstate. This problem can be solved if there is a dynamical mechanism that drastically reduces the dimensionality of the Hilbert space. In Hooft's proposal a gravitational type interaction originates the loss of information required for such reduction \cite{Blasone, Hooft}. However, there is not logical need for gravity from a formal point of view.

In DCRM there is a natural mechanism to bound from below the spectra of the quantum Hamiltonian $\widehat{H}(t,\tau,\hat{u},\hat{p})$ without introducing the gravitational interaction from the beginning \cite{Ricardo2005}. In particular, in the $U_t$ dynamics the classical Hamiltonian \eqref{randershamiltoniant} evolves towards an identically zero classical Hamiltonian. Thus, we impose on the  $U_t$ operator the quantum  constraint
\begin{align}
\lim_{t\to T} \widehat{H}(t,\tau,\hat{u},\hat{p})|\psi\rangle_t=0
\label{constraintontheHamiltonian}
\end{align}
 This constraint shows the emergent character of the local diffeomorphism invariance for the $U_t$ and $U_\tau$ dynamics when acting on an generic quantum state $|\psi\rangle\in\,\mathcal{H}$ in the  domain $t\to T$.

\section{Concentration of measure}
Under the hypothesis that in the domain $t\to T$ the $U_t$ is  a $1$-Lipschitz operator, we can find several interesting consequences. In particular, under this hypothesis,  {\it concentration of measure} as it appears in asymptotic theory of finite normed spaces  \cite{MilmanSchechtman2001}, metric geometry \cite{Gromov} and probability theory \cite{Talagrand} can be applied in DCRM. Let us describe very briefly what is concentration of measure. A measure metric space is a triplet $(\mathcal{T},\eta_P, \varepsilon)$ where $\mathcal{T}$ is a topological space, $\eta_P$ is a measure on $\mathcal{T}$ and $\varepsilon$ is a metric function on $\mathcal{T}$. Then the general property of concentration is usually quoted as follows \cite{MilmanSchechtman2001},
\begin{center}
 {\it In a measure, metric space, real $1$-Lipschitz functions of a large number of real variables are almost constant almost everywhere.}
\end{center}
The metric structure $\epsilon$ is necessary to define the $1$-Lipschitz condition; the measure structure $\eta_P$ is necessary to have a notion of almost everywhere.

The intuitive reason behind this phenomenon can be explained as follows in the case that $\mathcal{T}$ is a topological manifold (therefore, the notion of dimension of $\mathcal{T}$ is defined).  When the function $f:\mathcal{T}\to {\bf R}$ is under strict control, as it is the case of a $1$-Lipschitz function, the possibility that the difference between two values on the image is significant when they are found as a translation in a given direction is sharply cut off with the distance in the images between them. This is because when the number of variables is large, since similar differences will appear in other directions, the $1$-Lipschitz condition will be spoiled. They can be exceptions to these bounds, but the probability that this happens is zero. The principle of concentration is a bast generalization of the central limit theorem in probability theory.

 To illustrate how the concentration of measure arises in DCRM we assume that the internal dynamics operator $U_t$ is $1$-Lipschitz for values of the parameter $t$ close enough to $T$. This assumption is compatible with the fact that there is an equilibrium  limit is $1$-Lipschitz (by assumption). Then let us consider a $1$-Lipschitz function $f\in\,\mathcal{F}_D(T^*TM)$ of the form
  \begin{align*}
  {f}:\,\prod^N_{k=1}\, T^*TM^k_4\to R
  \end{align*}
with $N>>1$. Then one has that almost everywhere
\begin{align*}
{f}(u_1,...,u_N)& ={f}(v_1,...,v_N)+\,\mathcal{E}(N),
\end{align*}
where $\mathcal{E}(N)$ is a small error depending on the natural number $N$. This error $\mathcal{E}(N)$ quantified in function of $\epsilon(u,v)$ in the form of {\it concentration maps} \cite{MilmanSchechtman2001,Talagrand}. These concentration maps are usually exponential maps. The power of the concentration phenomenon will be exemplified in the following two sections.

\section{Example of concentration of measure and application to dynamical reduction of quantum states}

Let us consider an application of the concentration of measure in ${ R}^q$ with $q>>1$. $\eta_P$ is a measure and $f:{ R}^{q}\to { R}$ a real $1$-Lipschitz function. Then there is concentration given by (see for instance \cite{Talagrand} , pg. 8)
\begin{align}
\eta_P(|f-M_f|>\rho)\leq \, \frac{1}{2} \exp\Big(-\frac{\rho^2}{2\rho^2_P(f)}\Big),
\label{concentration2}
\end{align}
where we have adapted the example from \cite{Talagrand} to a Gaussian distribution $\eta_P$ with mean $M_f$  and standard deviation $\rho_P(f)$.  $\rho_P(f)$ has the physical interpretation of being the minimal resolution attainable when measuring the observable associated with the classical function $f:{ R}^{q}\to R$.

We apply this {\it example } of concentration to the function $f\in\,\mathcal{F}_D(T^*T{M})$. For  ${f}$ there is a maximal resolution $\rho_p(f)$ in their possible measurement outcome values.
In the $1$-Lipschitz dynamical regime of the evolution operator $U_t$, the function ${f}$ must be constant almost everywhere, since $f$ is $1$-Lipschitz in $(u,p)$ and $t$.
Moreover, for macroscopic observable effects one expect a relation of the type
 \begin{align}
\frac{\rho^2}{\rho^2_P(\tilde{f})}\simeq N^2,\quad 1<<N.
\label{comparingcomplexities1}
 \end{align}

 Let us restrict our considerations to the case when the quantum system corresponds to a fundamental particle  which quantum fields appear in the Standard Model. The natural number $N$ provides a measure of the {\it complexity} of the fundamental system compared with the complexity of the associated { quantum system}.
   The degree of complexity of a quantum state is of order $1$, since there is  one quantum particle involved compared with the degree of complexity of a DCRM, which is of order $N$. This order of complexity $1$ is of the same order than the dimension of the model space-time manifold $M_4$, the number of spin degrees of freedom, and other quantum numbers associated with the quantum mechanical description of a fundamental quantum particle.
 Let us consider the case $TM\cong R^{8N}$. If we make use of \eqref{comparingcomplexities1}, the concentration relation \eqref{concentration2} applied to the function ${f}$  in the $1$-Lipschitz dominated regime of $U_t$ becomes
  \begin{align}
  \eta_P\Big(|f-M_{f}|>\rho_P({f})\Big)\leq \, \frac{1}{2} \exp\Big(- 32\,{N^2} \Big).
  \label{concentration3}
  \end{align}
  Note that $M_{f}$ depends on the initial conditions $(u_1(0),...,u_{8N}(0),p_1(0),...,p_{8N}(0))$.
Since by assumption  $\rho_P(f)$ is small compared with $|f-M_f|$ and $N_0>>1$, there is concentration of measure around the mean $M_{f
}$. Thus, if a measurement of an observable associated with $f$ is performed, the result $M_{f}$ will be obtained with high certainty. This corresponds with a reduction of the quantum state. These reduction of the state  not only happens when the system is being {\it measured}, but it is an {\it spontaneous process}.
Such spontaneous processes happen even if there is not measurement. The driven force for this to happen is purely classical, since it appears in the limit $t\to T$ only (in the ergodic regime $\hat{H}\neq 0$ and therefore, it is not present). A measurement involves an additional process which is not classical and mediated by effective quantum gauge interactions between the system and a external quantum particle (for example, electromagnetic interaction). Such quantum process is amplified by the detector device.

\section{Concentration of measure and emergence of the weak equivalence principle in DCRM}
   We denote by {\it observable coordinate} $X^\mu(\mathcal{S}_i(\tau)$ associated to the system $\mathcal{S}_i,\,i\equiv \mathcal{S},A,B$ the coordinate of the system $\mathcal{S}_i$ measured by a macroscopic observer. In the concentration($1$-Lipschitz dynamical domain), such observable are $1$-Lipschitz. The observable coordinate does not depend on $t$, since emerge in the effective description of DCRM in the projection $(t,\tau)\to \tau$. Therefore, they correspond to measurements of position observable performed by a macroscopic observer.

    The configurations of the subsystems $A$ and $B$ can be represented in some special local coordinates on $TM$ by the coordinate representation
   \begin{align*}
   A\equiv (u_1(\tau),...,u_{N_A}(\tau),0,...,0) \quad \textrm{and}\quad B\equiv (0,...,0,v_1(\tau),...,v_{ N_B}(\tau)),\quad
   \end{align*}
   with $N=\,N_A+N_B,N_A,N_B>>1$ and $u(\tau)=\,u(t=T,\tau)$.
Locally, the full system $\mathcal{S}$ can be represented as
   \begin{align*}
   \mathcal{S}\equiv (u_1(\tau),...,u_{ N_A}(\tau),v_1(\tau),...,v_{ N_B}(\tau)).
   \end{align*}
By the concentration property \eqref{concentration2}, the evolution under the same initial conditions for the observable coordinates $X^\mu(\mathcal{S}(\tau))$, the configuration $X^\mu(A(\tau))$, the configuration $X^\mu(B(\tau))$ will differ after the evolution along the time $\tau$  such that
   \begin{align*}
  \eta_P\big(|X^\mu(\mathcal{S}_i(\tau))-M^\mu(\tau)|>\rho\big)_{t\to T}\sim C_1(i)\exp \big(-\,C_2(i) \frac{\rho^2}{2L^2_p}\big).
   \end{align*}
$C_2(i)$ is of order $32$.
For the $U_t$ dynamics it can happens that there is an interchange of fundamental degrees of freedom in the form of fundamental interactions with the environment. This can affect the motion of the center of mass $M^\mu(\mathcal{S}_i(t,\tau))$ in a rather intricate way, depending on the system $\mathcal{S}_i(\tau)$. However, if the system is in {\it free fall} in the sense of absence of interchange of matter with the environment, the center of mass coordinates $M^\mu(\mathcal{S}_i(\tau))$ follow a well defined ordinary differential equation,
 \begin{align}
 \frac{d}{d\tau} M^\mu(\tau)=\,{h}^\mu(\tau)
 \end{align}
 where the function ${h}^\mu$ is fixed by the equations of motion of the $k$-degrees of freedom by the properties of the measure $\mu_k$ and it does not depend on the system. We can assume that $\frac{d}{d\tau}\mu_k(t,\tau)=0$.
 Then the center of mass coordinates $M^\mu(\tau)$  will not depend on the system $\mathcal{S}$, $A$ or $B$.
 In this case, let us consider geometric configurations for the systems $\mathcal{S}_i,\,i=1,2,3$ such that the mean value  functions $\{M^\mu(\tau):R\to R, \tau\mapsto M^\mu(\tau)\}^4_{\mu=1}$ do not depend on the system $A$, $B$ or $\mathcal{S}$ at any time $\tau$. Since for macroscopic or quantum systems the quotient $\frac{\rho^2}{2L^2_p}\geq 32$,  there is concentration of the functions $\{X^\mu(\tau)\}^4_{\mu=1}$  around the same mean $\{M^\mu(\tau)\}^4_{\mu=1}$.

 In the limit $t\to T$ the Hamiltonian is constrained by the condition \eqref{constraintontheHamiltonian}. Such condition suggests that in the dynamical evolution described by the operator  $U_\tau$, the gravitational interaction must be included, since the dynamics is invariant under diffeomorphisms. Therefore, in the limit $t\to T$ we can decompose
 \begin{equation}
 \hat{H}_t=\,\hat{H}_{matter}+\,\hat{H}_{gravity},
 \end{equation}
 where $\hat{H}_{gravity}$ is the Hamiltonian responsible for the $U_t$ dynamics and must be introduced in order that the constraint \eqref{constraintontheHamiltonian} holds. In this interpretation, gravity is associated with the purely internal dynamics $U_t$ in the $t\to T$ limit. Moreover,
if we assume the value $N$, we can make a falsifiable prediction: in DCRM framework, the weak equivalence principle applied to the observable coordinates $\{X^\mu\}^4_{\mu=1}$ is exact up to a precision ${O}(\exp(-32 N^2)$ or higher. Thus, for $N=1$ the error should be of order $\exp(-32)$, while for $N=2$ it should be of order $\exp(-128)$.

 \section{Gravity as emergent interaction}

  If we collect all the characteristics of the $1$-Lipschitz interaction $U_t$ in the regime $t\to T$ we have that,
   \begin{enumerate}
   \item It is described by a theory invariant under infinitesimal diffeomorphism transformations, since the constraint \eqref{constraintontheHamiltonian} holds good,

   \item The weak equivalence principle for the center of mass functions $M^\mu(\mathcal{S}(\tau))$ holds good,

   \item There is a local maximal speed for the fundamental degrees of freedom of a DCRM and  local Lorentz invariance holds,

   \item It is a classical, macroscopic and universal interaction,

   \item It must be compatible with the existence of a maximal acceleration.
   \end{enumerate}
   Then the identification of a part of the $1$-Lipschitz interaction $U_t$ in the regime $t\to T$ and the gravitational contribution to the external interaction itself $U_\tau$ is a natural step, since the constraint \eqref{constraintontheHamiltonian} must hold. Also, let us remark  that the acceleration is universally bounded, in contrast with general relativity. Maximal acceleration appears because the local character of interactions, the existence of an universal minimal universal length, and the existence of an universal maximal speed for any dynamical degree of freedom. All these properties are present in DCRM.

The suggestion  that the gravitational interaction and objective reduction of the wave packet are related is not new (see for instance \cite{Adler2014, Diosi, Penrose, Smolin1986}).  However, we have argued that in the DRCM framework  classical gravity and the reduction of the wave packet are two aspects of the same concentration of measure phenomena. Moreover, our argument supports the classical and emergent nature of the gravitational interaction. From what we have said, the emergent nature is clear. The classical nature is irremediably associated to the emergence, since it is an interaction that appears significatively only in the regime where physical quantum and macroscopic systems are localized in position and speed variables.

\subsection*{Acknowledgements}This work was supported by PNPD-CAPES n. 2265/2011, Brazil and by ICTP, Italy.

\end{document}